# Thermal Design of Power Electronic Circuits

*R. Künzi*
Paul Scherrer Institute, Villigen, Switzerland

**Abstract**
The heart of every switched mode converter consists of several switching semiconductor elements. Due to their non-ideal behaviour there are ON state and switching losses heating up the silicon chip. That heat must effectively be transferred to the environment in order to prevent overheating or even destruction of the element. For a cost-effective design, the semiconductors should be operated close to their thermal limits. Unfortunately the chip temperature cannot be measured directly. Therefore a detailed understanding of how losses arise, including their quantitative estimation, is required. Furthermore, the heat paths to the environment must be understood in detail. This paper describes the main issues of loss generation and its transfer to the environment and how it can be estimated by the help of datasheets and/or experiments.

## 1 Introduction

Figure 1 shows the structure of a typical switched mode converter—in this case a buck converter—fed by a diode rectifier with input and output filters.

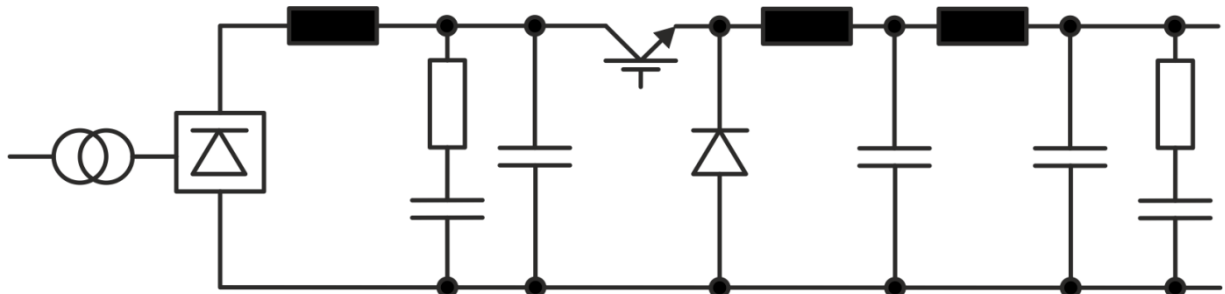

**Fig. 1:** Generic structure of a buck converter with diode rectifier and filters

Every component must be considered as a heat source. Also, capacitors carrying large alternating current (a.c.) heat up due to their ESR (equivalent series resistance), which can limit their lifetime, especially for electrolytic capacitors. Inductive devices like transformers and chokes have losses, and, due to their compact design, generate hot spots. Resistors are usually not critical as they are designated to dissipate power, and therefore designers are aware of their thermal behaviour.

The main heat sources are the semiconductors. They act as switching elements, i.e., they are either fully conducting or fully blocking. While the blocking losses are usually low enough to be negligible, the voltage drop in the ON state must be considered. At the transition between the ON and OFF states both the current through the semiconductor and the voltage across it are present for a short time causing switching losses.

Special care must be taken if the chip temperature and/or the base plate temperature oscillate. This phenomenon is known by the terms 'power cycling' and 'thermal cycling'. Temperature gradients cause additional mechanical stress to the soldering and welding joints inside a semiconductor module. The number of such thermal cycles is limited.

## 2 Heat sources

### 2.1 Resistors

Resistors with continuous load must be designed for a sufficient continuous heat transfer to the environment. At stable ambient temperature, a resistor's temperature is also stable. The power ratings given in datasheets are maximum ratings for very effective cooling and maximum temperatures. It is recommended to oversize the devices considerably in order to prevent hotspots or aging problems.

A resistor with a pulsed load must be designed to absorb the energy during the pulse. As a consequence of this the temperature of the resistor rises quickly. Pulse resistors therefore have a large amount of active material, such as wires or cast iron. After the pulse, there must be enough time for the active material to cool down, before the next pulse occurs. There must be adequate overload protection, i.e., limitation of the pulse duration, limitation of the pulse repetition frequency, or even a temperature measurement of the active material.

Cables must also be treated as resistors, and they must be designed and installed to allow sufficient heat transfer to the environment. Cables with sufficient cross-section should be used. This may result in higher costs at installation, but offsets the energy saving during the increased life period due to lower temperatures. The skin effect must also be considered: the penetration depth in copper is approximately 2.1 mm at 1 kHz and approximately 0.66 mm at 10 kHz. This reduces the effective cable cross-section and can lead to overheating. Large cable bundles should be avoided, when the current density is high, and there should be sufficient air circulation.

### 2.2 Magnetic components

Magnetic components like chokes and transformers dissipate power in both the core and the winding. Core losses are caused by *eddy current losses* and *hysteresis losses*.

Depending on the frequency, the eddy current losses are minimized by different means. For low frequencies up to approximately 1 kHz the cores are built with core plates: the thinner the core plates, the lower the eddy currents. For higher frequencies (from 1 kHz to approximately 10 kHz) powder cores are used. These can be seen as cores with extremely thin core plates, and therefore tend to have considerably fewer eddy current losses. In the frequency range above 10 kHz ferrite is the core material of choice. Ferrite cores have a very poor electric conductivity and therefore have almost no eddy currents.

Hysteresis losses arise from the periodic reversal of magnetism, which requires energy. The energy is proportional to the area of the hysteresis curve. Transformers are designed to have a small area of the hysteresis curve in order to minimize the losses. Be aware that the core losses of a transformer are also present at zero load! The hysteresis curves of d.c. chokes (which are designed to store a lot of magnetic energy) span a large area. The ripple current in such a choke causes the core to run through only a small part of the hysteresis area. The hysteresis losses depend very much on the amplitude and the frequency of the ripple current.

Winding losses, also called Cu losses, arise from the ohmic resistance of the winding. The dissipated power is approximately proportional to $I^2$. The resistance of a copper winding increases approximately 0.4% per kelvin, i.e., a temperature raise of 25 K increases the winding losses by approximately 10%.

There are good reasons to keep the temperatures of both the core and the winding as low as possible, even if it costs something: winding and core losses increase with higher temperatures. Losses in general are expensive; there are not only the costs of the electric energy, but also the costs for re-cooling. Lower temperatures increase the life period. This is especially important for components which are used permanently under high load.

## 2.3 Semiconductors

In most power electronic applications semiconductors are either switched on or off. Therefore, we have either current through the semiconductor or voltage across it, but never both at the same time. But that is only valid in an ideal world. In reality, there are voltage drops in the ON state, which cause conduction losses. There are also leakage currents in the OFF state, but they are usually very low and therefore not considered here. At the transition between the ON and OFF states both current through the semiconductor and voltage across it are present. This leads to high power dissipation for a very short time.

In order to estimate the losses in our application, we have to consult the datasheet. In the following the losses of a buck converter according to Fig. 1 are estimated. The converter is designed for a maximum output current $I_{Out}$ of 400 A and a maximum output voltage $V_{Out}$ of 200 V. The d.c.-link voltage $V_{dc}$ is 250 V. The switching frequency is 20 kHz. The combination of an IGBT (insulated-gate bipolar transistor) and a freewheeling diode is realized with a half bridge module FF600R06ME3 from Infineon. The graphs in Figs. 2–4 and Fig. 11 originate from its datasheet; see Ref. [1].

### 2.3.1 Conduction losses in the IGBT and the diode

During the ON state of the IGBT the output current flows through the IGBT. Its voltage drop can be derived from the output characteristic in the datasheet; see Fig. 2. The modulation index $m$ is the ratio $V_{Out}/V_{dc}$, and in a buck converter this is also the time ratio ON state/switching period. The voltage drop depends on the junction temperature. We make a pessimistic estimate for the junction temperature and get $V_{CE}$ from Fig. 2. In our example the conduction losses in the IGBT are given by

$$P_{cI} = m \cdot I_{Out} \cdot V_{CE} = 0.8 \cdot 200\,\text{A} \cdot 1.1\,\text{V} = 176\,\text{W} \,. \tag{1}$$

During the OFF state of the IGBT, the output current flows through the freewheeling diode. Its voltage drop can be derived from the forward characteristic figure in the datasheet; see Fig. 2. In contrary to the transistor, the diode usually has a negative temperature coefficient. Therefore we consider a rather low estimate for the junction temperature and get $V_F$ from Fig. 2. In our example the conduction losses in the diode are given by

$$P_{cD} = (1-m) \cdot I_{Out} \cdot V_F = (1-0.8) \cdot 200\,\text{A} \cdot 1.15\,\text{V} = 46\,\text{W} \,. \tag{2}$$

Note: If the converter operates at a low modulation index for the maximum output current, the diode has many more conduction losses!

Some manufacturers specify $V_{CE}$ and $V_F$ only at the chip level, as this is relevant to determine the losses in the silicon chip. In addition to this there is also a resistance between the silicon chip and the terminals. These losses are mainly transported to the environment through the terminals and the connected cables. They may also be considered for an overall estimate of the losses. In the datasheet [1] this resistance is called $R_{CC'+EE'}$ and its value is 1.1 mΩ per switch. In our example with $I_{Out}$ = 200 A and $m$ = 0.8 this leads to an additional voltage drop of 0.22 V in each branch and to additional losses of 35 W in the IGBT branch and 9 W in the diode branch.

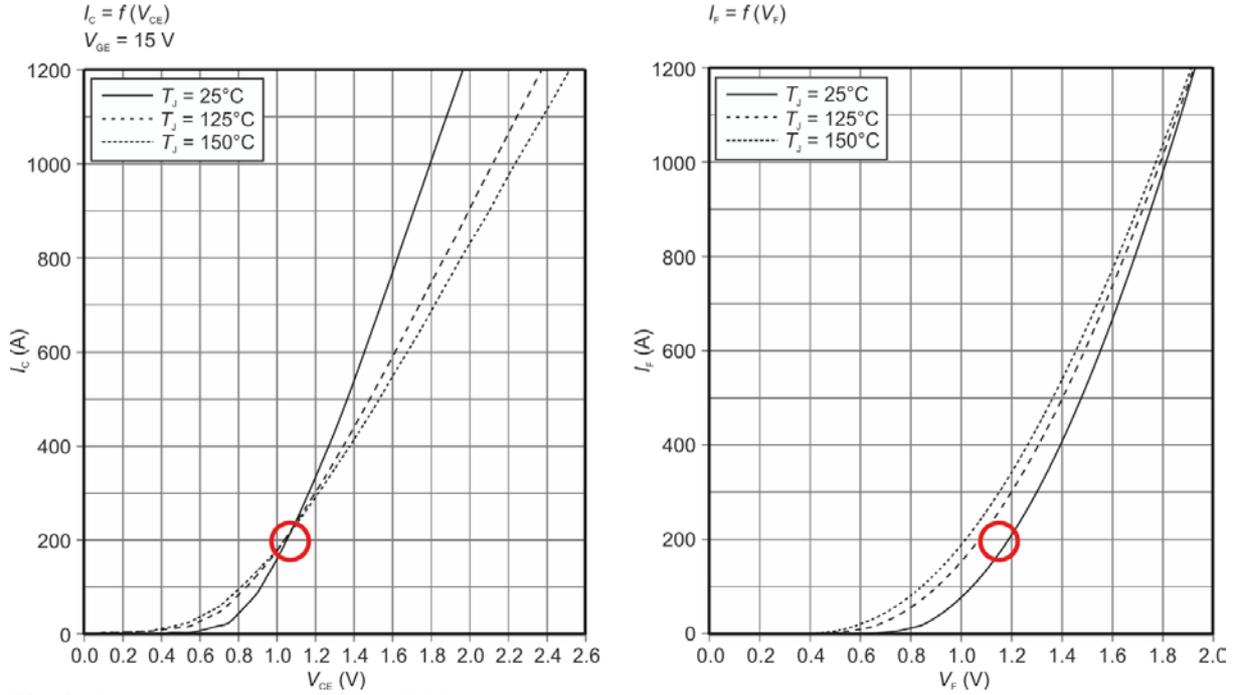

**Fig. 2:** Output characteristic of the IGBT for a gate-emitter voltage $V_{GE}$ of 15 V (left) and forward characteristic of the diode (right).

### 2.3.2  Switching losses in the IGBT

The dissipated energy $E_{on}$ and $E_{off}$ at each switching is given in the datasheet. The graphs are given for discrete reference junction temperatures $T_{Ref}$, e.g., 125ºC or 150ºC; see Fig. 3. For a different junction temperature $T_J$ a correction factor applies; refer to Eq. (3). Accordingly, the graphs are given for a reference blocking voltage $V_{ref}$, e.g., 300 V. For a different blocking voltage $V_{CE}$, a correction factor applies; refer to Eq. (3). The switching losses in the IGBT are given by

$$P_{sI} = f_s \cdot (E_{on} + E_{off}) \cdot \left(1 + TC(T_J - T_{Ref})\right) \cdot (V_{CE}/V_{ref})^{K_V} . \tag{3}$$

For transistors, $TC \sim 0.003$ K$^{-1}$ and $K_V \sim 1.3$–1.4; see Ref. [2], p. 287. With the values from our example with $f_s = 20$ kHz, $V_{CE} = V_{dc} = 250$ V, and an estimated junction temperature $T_J$ of 90ºC, we get

$$P_{sI} = 20{,}000 \text{ s}^{-1} \cdot (5+8) \cdot 10^{-3} \text{ J} \cdot (1 + 0.003 \cdot (90-125) \text{ K}) \cdot \left(\frac{250 \text{ V}}{300 \text{ V}}\right)^{1.35} = 182 \text{ W}.$$

Be aware that the switching losses are highly dependent on the gate resistance; see Fig. 3 (right). Hard switched converters are usually optimized for very fast switching in order to minimize the losses. However, if the switching needs to be softer, e.g., for EMC reasons (electromagnetic compatibility), this can be achieved by increasing the gate resistor. As shown in Fig. 3 on the right, this dramatically increases the switching losses.

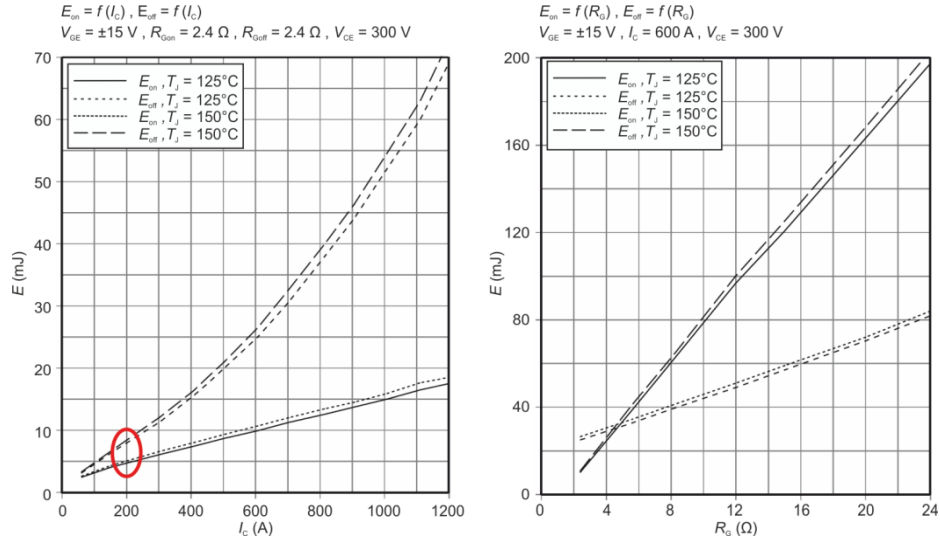

**Fig. 3:** Switching losses in the IGBT for a gate-emitter voltage $V_{GE}$ of ±15 V

### 2.3.3 Switching losses in the diode

The dissipated energy $E_{rec}$ at each switching OFF is given in the datasheet. The graphs are given for discrete reference junction temperatures $T_{Ref}$, e.g., 125ºC or 150ºC; see Fig. 4. For a different junction temperature $T_J$ a correction factor applies; refer to Eq. (4). Accordingly, the graphs are given for a reference blocking voltage $V_{ref}$, e.g., 300 V. For a different blocking voltage $V_{CE}$, a correction factor applies; refer to Eq. (4). The switching losses in the diode are given by

$$P_{sD} = f_s \cdot E_{rec} \cdot \left(1 + TC(T_J - T_{Ref})\right) \cdot (V_{CE}/V_{ref})^{K_V}. \tag{4}$$

For diodes, $TC \sim 0.006$ K$^{-1}$ and $K_V \sim 0.6$; see Ref. [2], p. 287. With the values from our example with $f_s$ = 20 kHz, $V_{CE} = V_{dc}$ = 250 V, and an estimated junction temperature $T_J$ of 90ºC, we get

$$P_{sD} = 20{,}000 \text{ s}^{-1} \cdot 5 \cdot 10^{-3} \text{ J} \cdot (1 + 0.006 \cdot (90 - 125) \text{ K}) \cdot \left(\frac{250 \text{ V}}{300 \text{ V}}\right)^{0.6} = 71 \text{ W}.$$

In contrary to the IGBT, a softer switching (higher gate resistance $R_G$) reduces the switching losses of the freewheeling diode remarkably; see Fig. 4 (right).

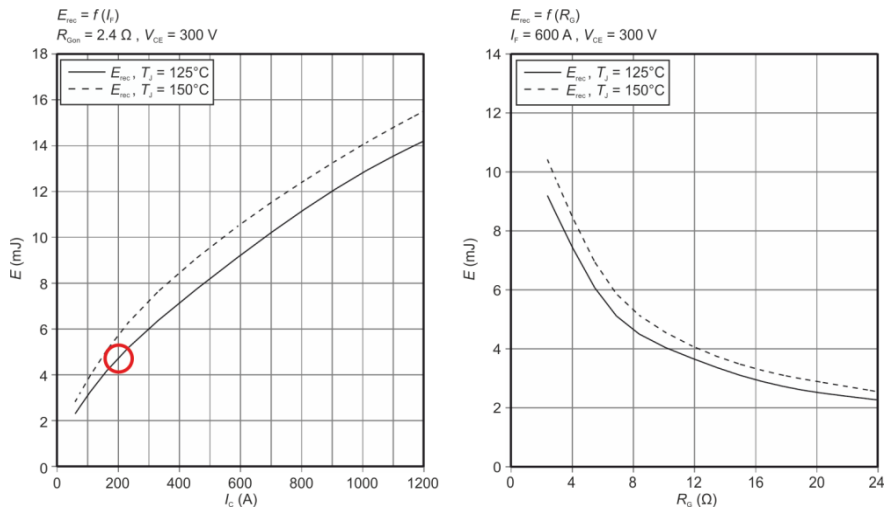

**Fig. 4:** Switching losses in the diode

*2.3.4    Total IGBT and diode losses*

Inside the module the heat is generated in two different silicon chips, the IGBT and the diode. The IGBT losses $P_\text{I}$ and the diode losses $P_\text{D}$ are given by

$$P_\text{I} = P_\text{cI} + P_\text{sI} = 176\text{ W} + 182\text{ W} = 358\text{ W}, \tag{5}$$

$$P_\text{D} = P_\text{cD} + P_\text{sD} = 46\text{ W} + 71\text{ W} = 117\text{ W}. \tag{6}$$

They are needed to determine the junction (chip) temperatures; see Section 3.2. In addition to that the module lead resistance $R_{\text{CC'+EE'}}$ causes terminal losses $P_\text{Term}$. The overall losses for the entire module $P_\text{M}$ are given by

$$P_\text{M} = P_\text{I} + P_\text{D} + P_\text{Term} = 358\text{ W} + 117\text{ W} + 44\text{ W} = 519\text{ W}. \tag{7}$$

This figure is needed to determine the overall converter losses and the efficiency.

## 3    Heat transfer

### 3.1    Packages

The losses as described in Section 2 originate in the silicon chips and must be transferred to the environment by means of heat sinks. Different packaging concepts are used (see Fig. 5) depending on the power rating.

For small discrete elements mounted on a PCB, the heat flows via the electrical connections and/or the air. The tracks on the PCB can be used as heat conductors in order to spread the heat to a large area or to lead the heat to a heat sink or the housing. See Section 3.5 for a practical example. The calculation of the resulting junction temperatures is difficult, and requires a detailed thermal modelling of the entire arrangement (semiconductor and PCB) and a time-consuming finite element analysis. Therefore usually a 'trial and error' approach is used.

Larger discrete elements have a cooling surface, which can be soldered to a copper pad or pressed onto a heat sink. Refer to Section 3.5 for a practical example. The thermal resistance between the silicon chip and the cooling surface is given in the datasheet. Note that usually the cooling surface is connected to one of the electrical connections. If isolation is needed, it needs to be realized separately.

Large semiconductor modules have isolated cooling plates, which can be connected to ground level. Refer to the example in Section 2.3. Most of the losses produced in such elements are dissipated through this cooling plate.

Very large semiconductors are packed in press-pack housings. They have large contact surfaces, which serve as both electrical and thermal connections.

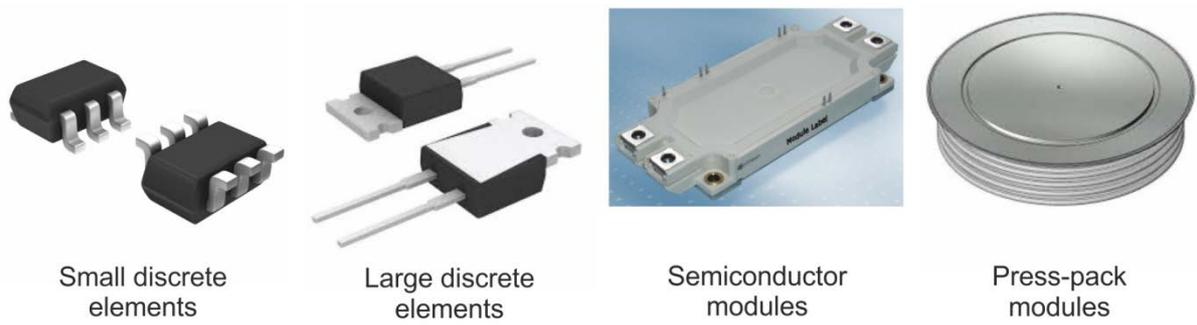

**Fig. 5:** Different packages for semiconductor devices

Figure 6 shows typical cross-sections of an isolated module package and a press-pack module. In the isolated module package the heat from the silicon chips must flow through the ceramic insulator and the soldering joints to the copper base. The press-pack housing does without soldering joints and ceramic insulator, and therefore has a much more effective heat transfer to the heat sink. Furthermore it can be cooled from both sides, but the two heat sinks are on different electrical potentials, usually neither of them connected to ground.

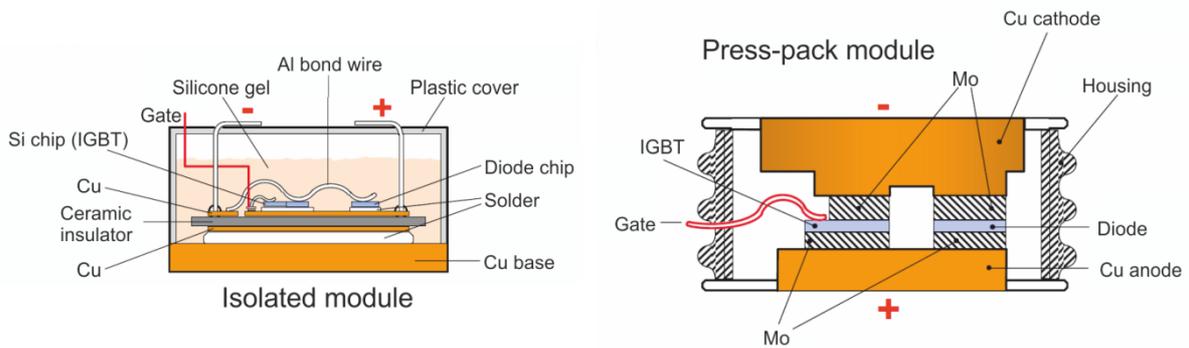

**Fig. 6:** Typical cross-sections for different housing concepts

### 3.2    Steady-state temperatures

The losses in both the IGBT and the freewheeling diode cause temperature differences in the thermal resistor network, as shown in Fig. 7. In steady state, the losses and the ambient temperature are constant. Therefore the resulting temperatures in the module are also constant.

The thermal resistances $R_{thJC}$ (junction to case) for both the IGBT and the diode can be found in the datasheet. Note: In data sheets the Cu base of an isolated module is also called 'case'. The thermal resistance $R_{thCH}$ (case to heat sink) is also given in the datasheet; this depends not only on the module itself, but also on the way the element is mounted to the heat sink. The surfaces of the device and the heat sink both have a certain roughness and therefore there are inherent air gaps between the two. In order to achieve an effective heat flow, these air gaps must be filled by a Thermal Interface Material (TIM). The application notes from the semiconductor suppliers for proper mounting of the element should be consulted; an example is given in Ref. [3]. The thermal resistance of the heat sink $R_{thHA}$ (heat sink to ambient) is either given by the manufacturer or can be determined by means of an experiment.

With this information, the corresponding temperature differences can be calculated by multiplying the power flow by the thermal resistance.

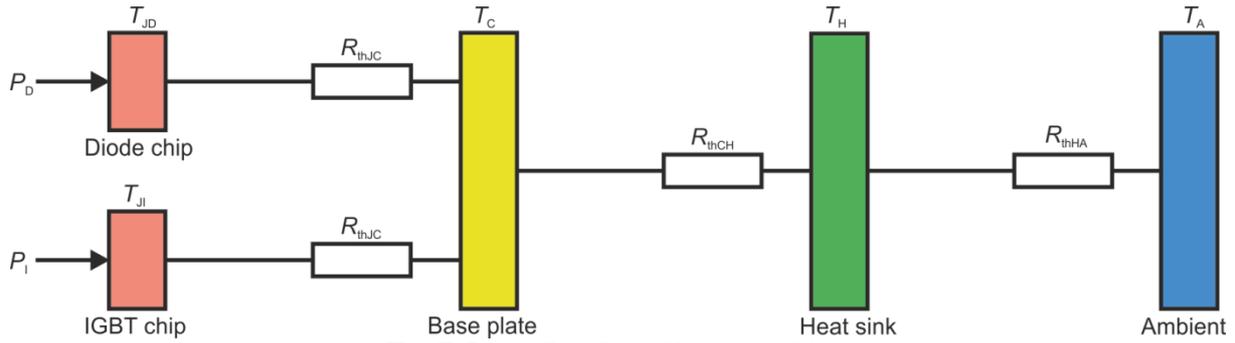
**Fig. 7:** Power flow from silicon to ambient

For the example in Section 2.3 we found losses of 358 W for $P_I$ and 117 W for $P_D$. In the datasheet [1] we find thermal resistances of 0.09 K/W for $R_{thJC}$ for the IGBT, 0.15 K/W for $R_{thJC}$ for the diode, and 0.009 K/W for $R_{thCH}$. The thermal resistance of the heat sink $R_{thHA}$ we assume to be 0.1 K/W. These figures are recorded in Fig. 8.

On multiplying the power flow by the corresponding thermal resistances we get the temperature differences $\Delta T_{JCI}$ (IGBT junction to case) = 32 K, $\Delta T_{JCD}$ (diode junction to case) = 18 K, $\Delta T_{CH}$ (case to heat sink) = 4 K, and $\Delta T_{HA}$ (heat sink to ambient) = 48 K. If we assume the ambient temperature to be 35ºC we get the absolute temperatures $T_H$ (heat sink) = 83ºC, $T_C$ (base plate) = 87ºC, $T_{JD}$ (diode junction) = 105ºC, and $T_{JI}$ (IGBT junction) = 119ºC. The results are summarized in Fig. 8.

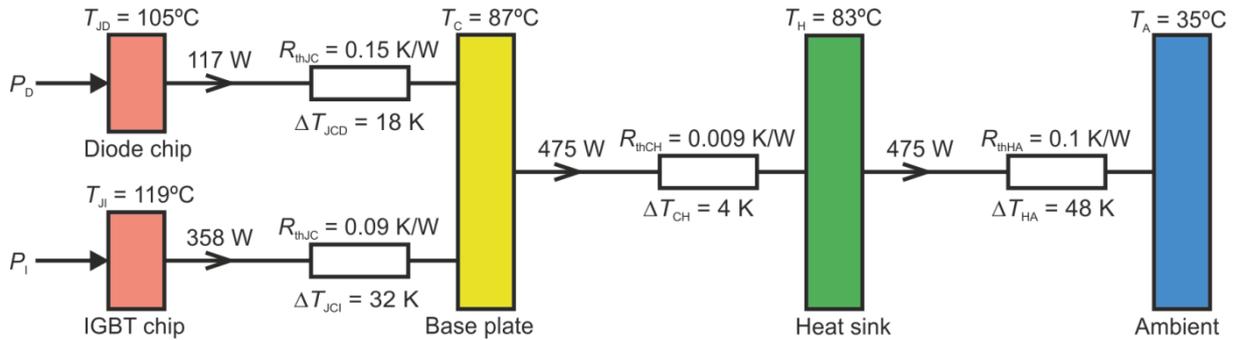
**Fig. 8:** Power flow for example in Section 2.3

According to the datasheet [1], the IGBT module used for our example may be operated at junction temperatures up to 150ºC. The loss calculation and the thermal calculations are not very accurate, and we also have to consider long-term drifts of the parameters involved. Therefore a safety margin of at least 25 K is necessary. Our example revealed a margin of 31 K, which is quite close to that limit. In general, lower temperatures raise the MTBF (mean time between failures) and the long-term reliability. It is left to the designer to make a reasonable trade-off between exploiting the elements and long-term reliability.

### 3.3    Transient thermal impedance

The loss and temperature calculations mentioned in Section 3.2 are valid for steady-state conditions only. If the losses vary with time (pulsating or oscillating), the thermal impedance has to be considered instead of the simple thermal resistance. From the thermal point of view, the arrangement shown in Fig. 6 (left) can be seen as a row of thermal capacitances (Si chip–solder–ceramic insulator–solder–Cu base) interconnected by thermal resistances. This leads to the continued fraction model, also known as the Cauer model, the T-model, or the ladder network (see Fig. 9). This model requires detailed knowledge of the material characteristics of the individual layers. However, the correct modelling of the thermal spreading is difficult.

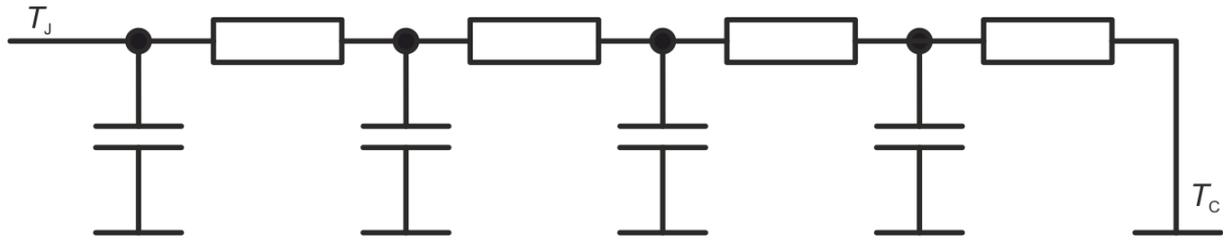
**Fig. 9:** Cauer model

Most manufacturers publish data for the partial fraction model, also known as Foster model or the Pi model (see Fig. 10).

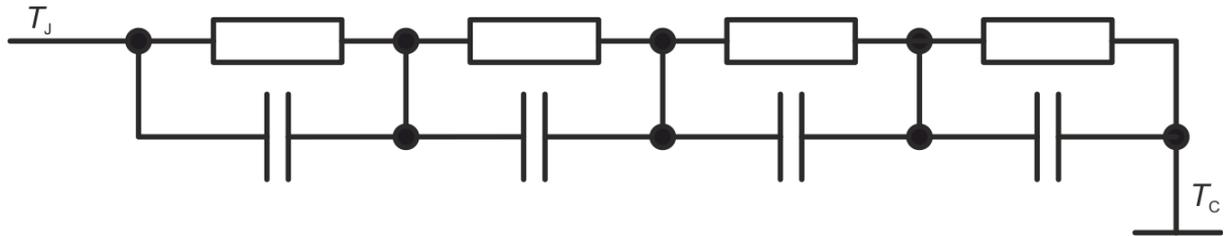
**Fig. 10:** Foster model

This model does not represent the layer sequence; the network nodes do not have any physical correlation. The thermal impedance

$$Z_{\text{thJC}}(t) = \sum_{i=1}^{n} r_i \cdot \left(1 - e^{-\frac{t}{\tau_i}}\right) \tag{8}$$

as a function of time is usually given in the datasheet; see Fig. 11 and Ref. [1]. The sum of the individual thermal resistances $r_i$ corresponds to the steady-state thermal resistance $R_{\text{thJC}}$. The time constants are typically in the region of 10 to 100 ms. If the power fluctuations are faster than that, e.g., losses during one switching period, we may simplify the calculation and consider the average power only, as in Section 2.3.

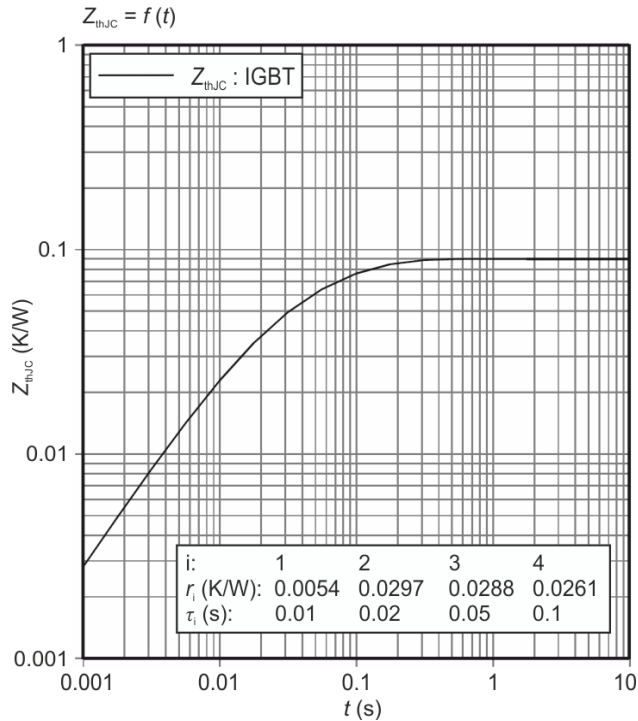
**Fig. 11:** Transient thermal impedance

Once the losses and the base plate temperature as a function of time are known, the junction temperature as a function of time can be determined as

$$T_{\text{J}}(t) = P(t) \cdot Z_{\text{thJC}}(t) + T_{\text{C}}(t). \tag{9}$$

For more information about transient thermal impedance see Ref. [4]; note that Figs. 9 and 10 originate from this reference.

### 3.4 Temperature oscillations

When the junction and the case temperatures oscillate, material fatigue has to be considered. Problems arise from different expansion coefficients for different materials and conjoined mechanical stress; see Table 1 and Fig. 6.

Table 1: Expansion coefficients for different materials

| Material | Expansion coefficient ($10^{-6}$/K) |
|---|---|
| Silicon | 4.1 |
| Copper | 17 |
| Aluminium | 24 |
| Molybdenum | 5 |
| Solder | 15–30 |
| Ceramic | 5–9 |

#### 3.4.1 Thermal cycling

Temperature oscillations of the base plate are referred to as 'thermal cycling'. There is a large-area soldering joint between the ceramic insulator and the base plate, which accomplishes a good thermal conductivity. The expansion coefficients for copper and solder material match reasonably well, but the ceramic material differs by a factor of 2 or more. This causes mechanical stress to the soldering joint, which leads to accelerated aging. Finally, the thermal resistance increases, the temperature oscillations become even larger, and the device fails.

Thermal cycling problems can be avoided by using water cooling with a constant water inlet temperature, which keeps the temperature of the heat sink and the base plate nearly constant.

#### 3.4.2 Power cycling

Temperature oscillations of the silicon chip are known as 'power cycling'. The expansion coefficients of aluminium (bond wires) and silicon (semiconductor chip) differ by a factor of 6. When the temperature oscillates there is mechanical stress to the bond-wire welding. After some time, the bond wires start to lift off. This causes an increased voltage drop $V_{\text{CEsat}}$, which increases the losses. This leads to an even larger temperature oscillation, and the device finally fails.

Press-pack devices do not have bond wires for the main connections and are therefore less sensitive to power-cycling stress.

In the mid-1990s in the scope of the LESIT project, standard modules with base plates from different manufacturers were tested regarding their power-cycling capability. During the project, a large volume of data was collected. Figure 12 shows the results for different temperature oscillations at three medium temperatures $T_{\text{m}}$.

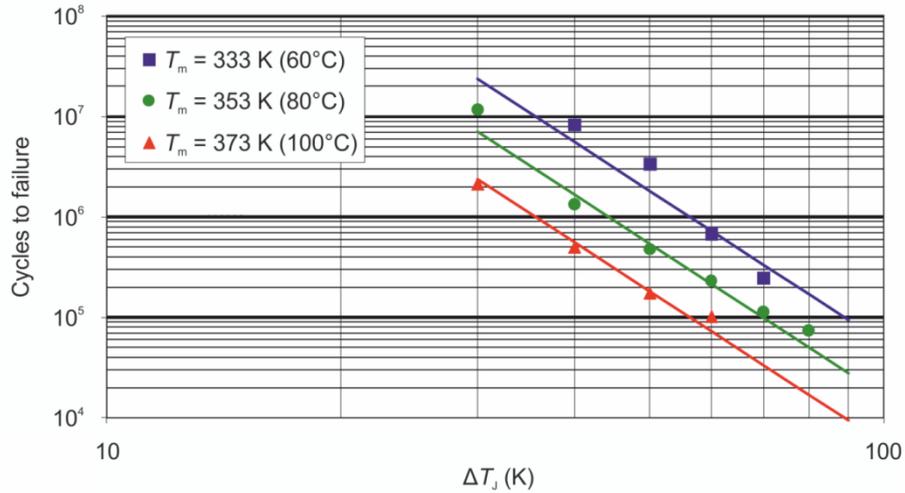

**Fig. 12:** Results from the LESIT project

Almost 20 years have passed, and the development of better packaging has evolved. Today there are specialized and improved devices on the market. However, for a first assessment, the LESIT figures are still valuable.

### 3.4.3    *Practical experience with power cycling*

The output current of the booster dipole power supply (PS) of the Swiss Light Source (SLS) oscillates with a 320 ms cycle time, i.e., with a frequency of 3.125 Hz, from zero to full current and back to zero. The SLS works in top-up mode, i.e., a short refilling of the storage ring takes place approximately once every 100 s. In the first 4 years of operation the booster PS was permanently on. Since then, the PS is only turned on during the refilling procedure of the storage ring, in order to save energy. During the remaining time, it is on standby. After 3.5 years in this 'energy-saving mode' we experienced two IGBT failures within a month. That was a clear sign of a systematic and reproducible aging problem, probably due to power cycling.

Measurements combined with simulations showed a temperature excursion as shown in Fig. 13. During standby, the chip temperature corresponds to the cooling-water inlet temperature, which is stabilized to 30°C. Once every 100 s the chip temperature is raised from 30°C to 86°C. For approximately 5 s there is a second oscillation with a 320 ms cycle time superimposed, where the chip temperature oscillates between 68°C and 82°C. This raised the question: which of the two oscillations killed the IGBTs?

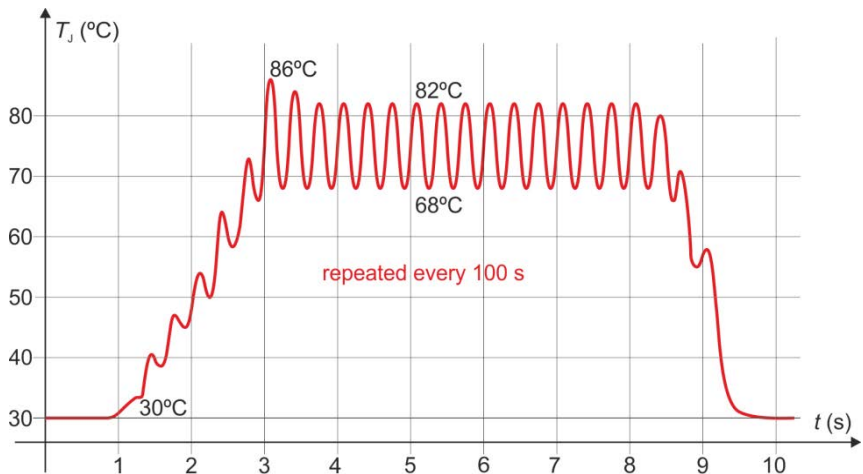

**Fig. 13:** Practical example of an IGBT junction temperature excursion

The 320 ms oscillation has a medium temperature of 75ºC and a peak-to-peak amplitude of 14 K. By extrapolating the LESIT results (see Fig. 14, dotted line) we can expect a lifetime of approximately $5 \times 10^8$ cycles. During the first 4 years of continuous operation (40 weeks per year) we gathered approximately $3.0 \times 10^8$ cycles. During the following 3.5 years in energy-saving mode (40 weeks per year, 16 cycles every 100 s) we gathered another $1.5 \times 10^7$ cycles, which sums to a total of $3.2 \times 10^8$ cycles. This is considerably below the expected lifetime. Furthermore, it is doubtful whether such an extreme extrapolation of the LESIT results is appropriate.

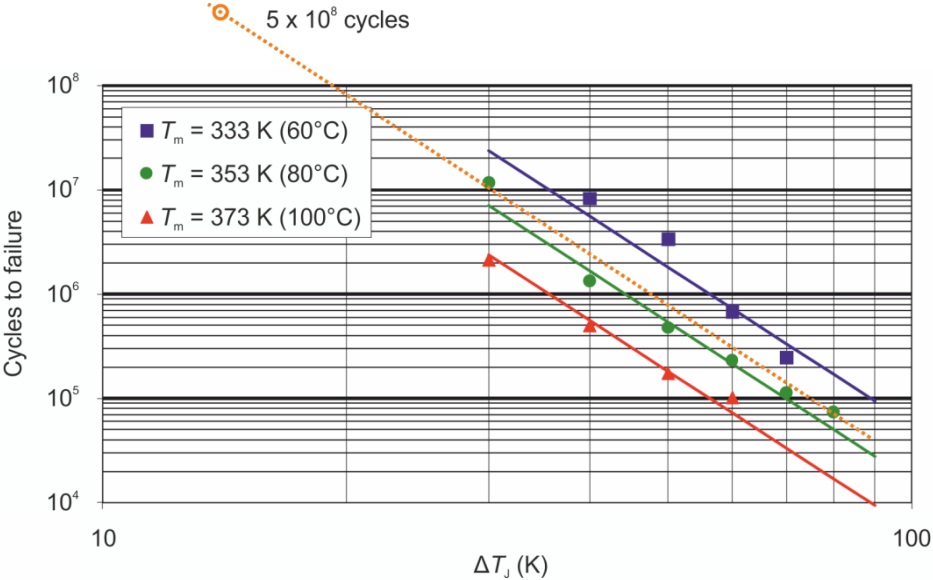

**Fig. 14:** Extrapolation of the LESIT results for the 320 ms cycles

The 100 s oscillation has a medium temperature of 58ºC and a peak-to-peak amplitude of 56 K. By interpolating the LESIT results (see Fig. 15, dotted line) we can expect a lifetime of approximately $1 \times 10^6$ cycles. During the first 4 years of continuous operation, there were only a few such cycles. During the following 3.5 years in energy-saving mode (40 weeks per year, one cycle every 100 s) we gathered approximately $8.5 \times 10^5$ cycles. This is very close to the expected lifetime and most likely killed the IGBTs.

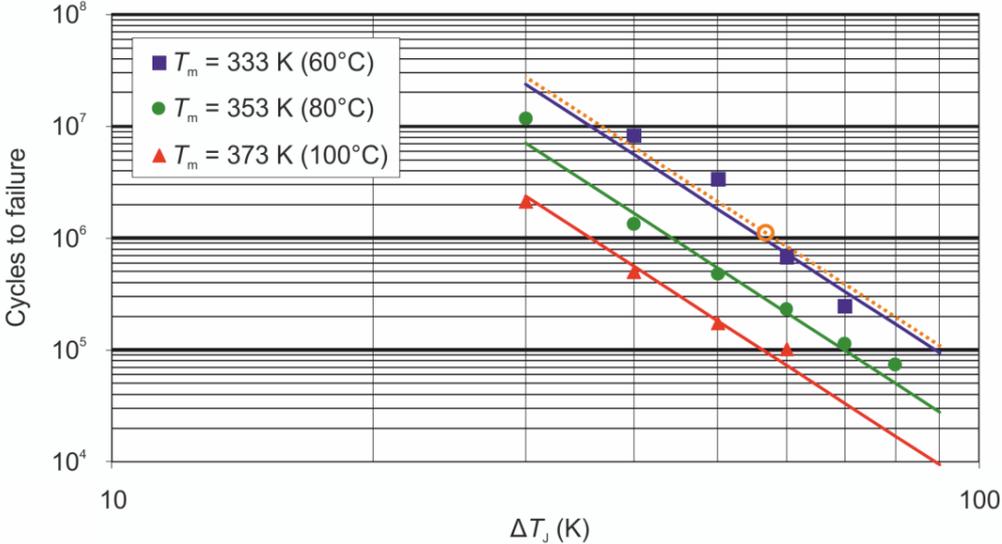

**Fig. 15:** Interpolation of the LESIT results for the 100 s cycles

## 3.5 Practical example: thermal design on a PCB

For the free electron laser SwissFEL at the Paul Scherrer Institute (PSI), a new bipolar 10 A/24 V converter has been developed, which is installed on a 100 mm × 160 mm PCB. The calculation of the resulting junction temperatures is difficult; this would require a detailed thermal modelling of the entire arrangement (semiconductor and PCB) and a time-consuming finite element analysis. Instead, we designed a first prototype, measured the temperature distribution on the board, and made improvements to the design. Figure 16 depicts a thermal image of the first prototype. It shows a very inhomogeneous board temperature with several hot spots up to 117ºC.

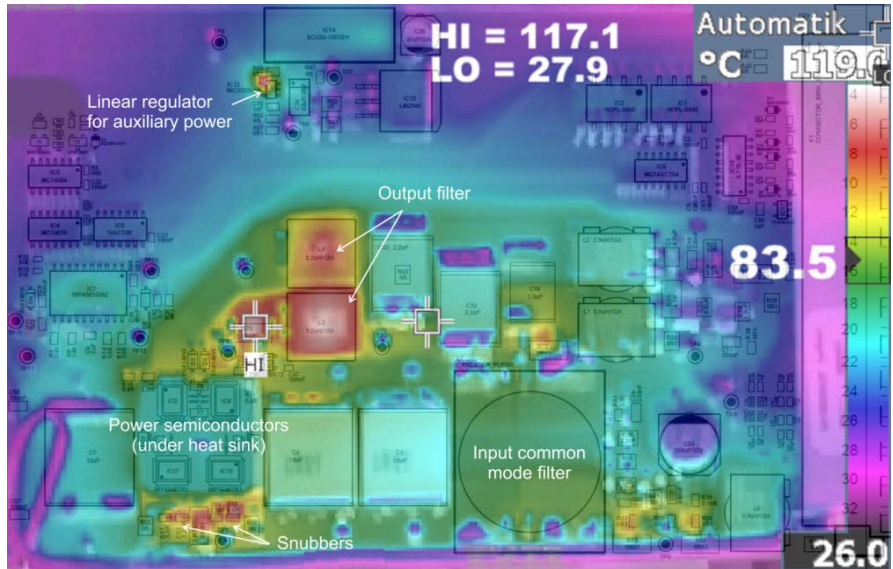

**Fig. 16:** Thermal image of the first prototype

For EMC reasons the entire PCB must be enclosed in a metallic housing. It was one of the main goals of the development project to design a converter that does not need any forced cooling. This requires a cooling design that uses the metallic shielding for a heat sink. In order to transfer the heat from the various hot spots on the PCB to the shielding, improvements in the design were necessary.

The thermal conductivity for materials used on a PCB differs greatly; see Table 2. The PCB core material is only about 10 times better than air, but 1000 times worse than metals. Heat-transfer foils are rather bad but unavoidable because they perform the electrical isolation. Therefore the strategy must be to conduct the heat as much as possible through metals.

**Table 2:** Thermal conductivity $\lambda$ for different materials

| Material | Thermal conductivity $\lambda$ (W/K m) |
| --- | --- |
| Gold | 318 |
| Silver | 429 |
| Copper | 401 |
| Aluminium | 237 |
| Steel | 50 |
| Heat-transfer foil | 2 |
| PCB core material (FR-4) | 0.3 |
| Air | 0.025 |

The four MOSFETs for the converter are mounted on the top side of the PCB on an area of approximately 2 cm × 2 cm = 4 cm². At full load the losses are approximately 10 W. The core thickness is 1.6 mm. The thermal resistance from top to bottom through the PCB core material is

$$R_{\text{th}} = \frac{l}{\lambda \cdot A} = \frac{1.6 \cdot 10^{-3} \text{ m}}{0.3 \frac{\text{W}}{\text{K} \cdot \text{m}} \cdot 4 \cdot 10^{-4} \text{ m}^2} = 13.3 \text{ K/W}. \tag{10}$$

This causes a temperature difference from top to bottom of

$$\Delta T = R_{\text{th}} \cdot P = 13.3 \frac{\text{K}}{\text{W}} \cdot 10 \text{ W} = 133 \text{ K}. \tag{11}$$

These figures clearly show that the heat transfer through the PCB core material is not sufficient.

Vias (borings coated with copper) are used for electrical connections from one layer to another. Such vias (see Fig. 17 for the geometry) can also be used to transfer the heat from the top layer to the bottom one.

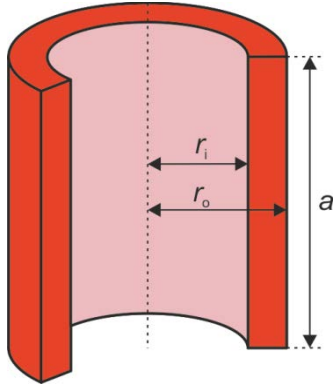

**Fig. 17:** Geometry of a via

The copper coating inside the boring has an outside radius $r_o$ of 0.175 mm and an inside radius $r_i$ of 0.15 mm. Its length $a$ is 1.6 mm, which corresponds to the core thickness of the PCB. The effective cross-section is

$$A = (r_o^2 - r_i^2) \cdot \pi = (0.175^2 - 0.15^2) \text{ mm}^2 \cdot \pi = 0.0255 \text{ mm}^2. \tag{12}$$

The resulting thermal resistance for a single via is therefore

$$R_{\text{th}} = \frac{l}{\lambda \cdot A} = \frac{1.6 \cdot 10^{-3} \text{ m}}{401 \frac{\text{W}}{\text{K} \cdot \text{m}} \cdot 0.0255 \cdot 10^{-6} \text{ m}^2} = 157 \text{ K/W}. \tag{13}$$

On an area of 4 cm² there is space for 8 × 8 = 64 parallel vias, which reduces the overall thermal resistance to 2.45 K/W. This thermal resistance is in parallel to the one of the PCB core materials, and the resulting temperature difference

$$\Delta T = R_{\text{th}} \cdot P = \frac{1}{\frac{1}{13.3 \frac{\text{K}}{\text{W}}} + \frac{1}{2.45 \frac{\text{K}}{\text{W}}}} \cdot 10 \text{ W} = 20.7 \text{ K} \tag{14}$$

between the top and bottom layers is acceptable.

Originally we had only one 70 µm layer for the signal and power wiring (Fig. 18, left). In the final design (Fig. 18, right), there are 105 µm layers on the top and bottom sides and two additional

35 µm layers in between. All four layers are electrically and thermally connected together. That reduces the current density in the power wiring and allows for an effective heat spread and heat transfer from top to bottom. The overall thickness of the PCB was reduced from 1.7 mm to 1.4 mm.

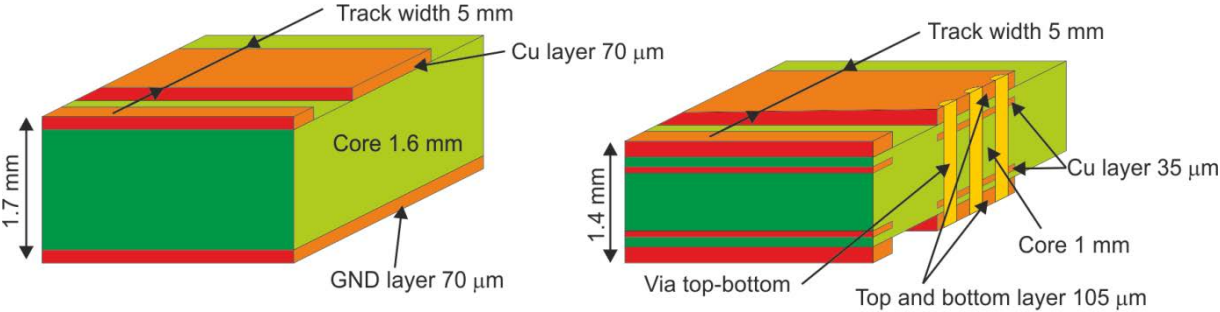

**Fig. 18:** Original (left) and final (right) PCB structure

Figure 19 shows the cross-section of the whole arrangement at the main heat source (the MOSFET switches) and its corresponding thermal model. The MOSFETs have cooling plates on top of them, onto which a heat sink is bolted. These cooling plates and the bottom layer of the PCB are electrically isolated by means of heat-transfer foil. The heat is transferred to the environment along three parallel paths: through the PCB and the aluminium bar, through the heat sink, and through the mounting bolt.

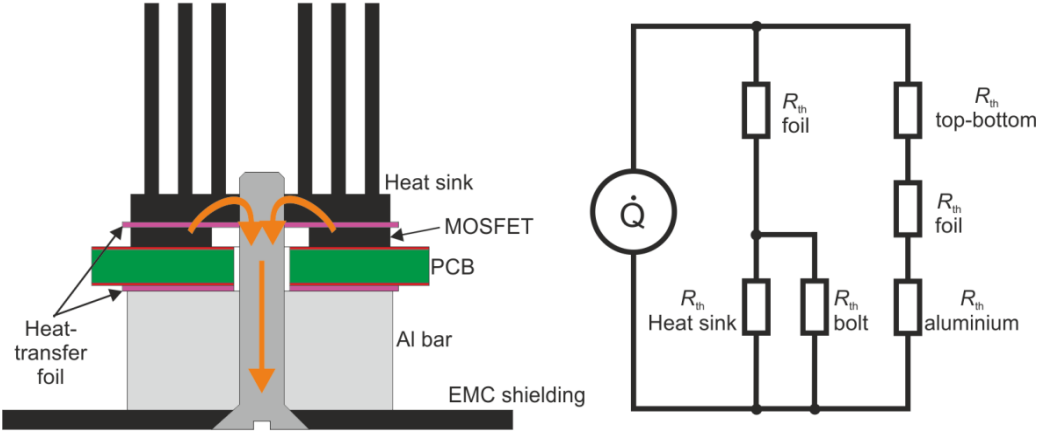

**Fig. 19:** PCB cross-section (left) and thermal model (right)

Figure 20 shows the bottom side of the final PCB. In total three aluminium bars transfer the heat to the EMC shielding. Between the aluminium bars and the Cu pads on the PCB there is an electrically isolating heat-transfer foil.

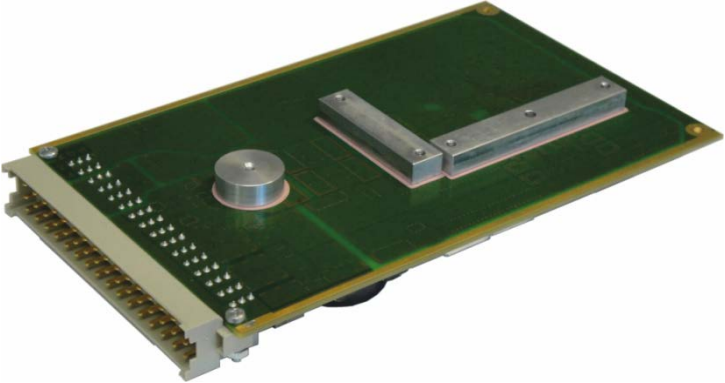

**Fig. 20:** Bottom side of the PCB

As seen in Fig. 16, there are several components that generate hot spots. Four components were exchanged in order to eliminate these hotspots. Figure 21 shows the location of these components on the PCB. Refer also to the thermal images in Figs. 16 and 22.

(1) the auxiliary d.c./d.c. converter 24 V/15 V in THT (through hole technology) in a plastic case was replaced by a SMD (surface mount device) type in a metallic case with better thermal behaviour;

(2) two chokes of 8.2 $\mu$H/13 A each were replaced by four chokes of 3.9 $\mu$H/12 A each, such that the heat source area could be enlarged;

(3) the linear auxiliary d.c./d.c. converter 15 V/5 V was replaced by a switched mode converter with much fewer losses;

(4) the 100 mW snubber resistors were replaced by 2 W types, which are larger and offer better heat transfer.

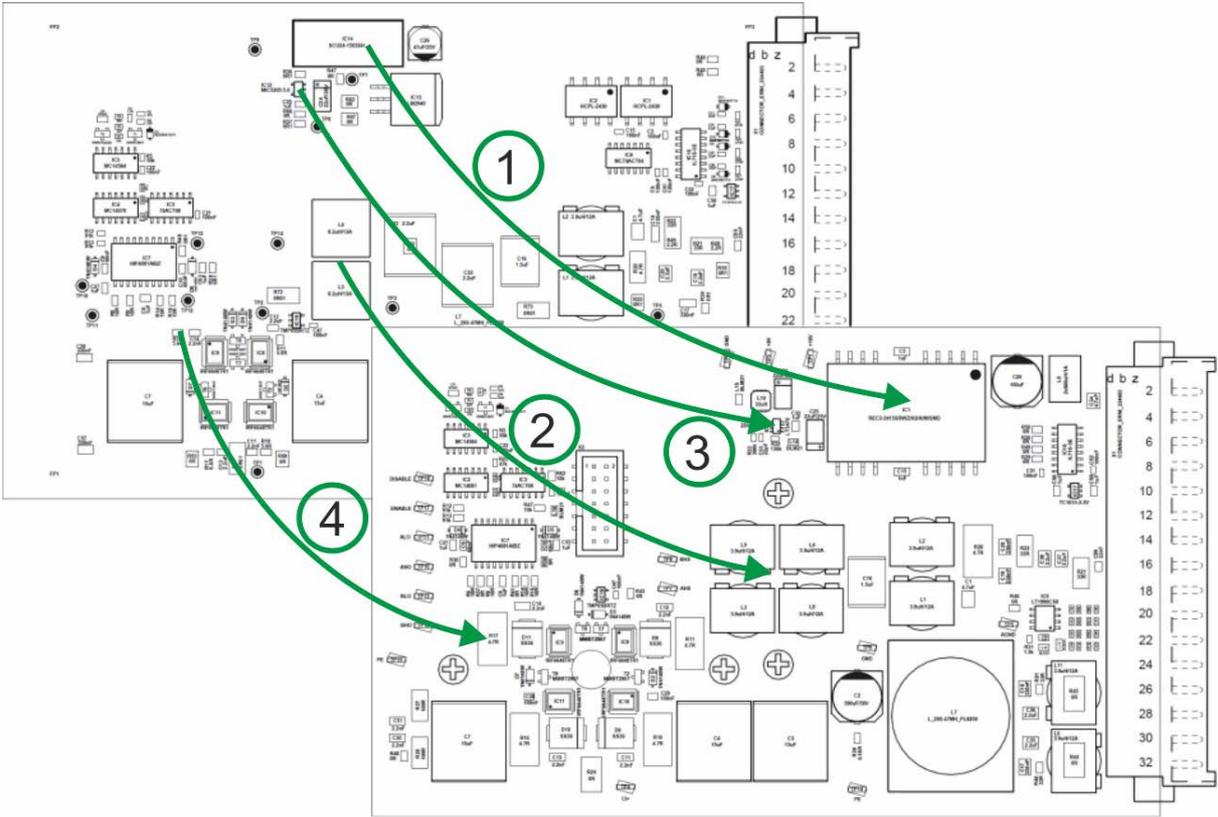

**Fig. 21:** Replacement of components that cause hotspots

Figure 22 shows the thermal image of the final design with the same temperature scaling as for Fig. 16. It shows a much more homogeneous temperature distribution. The majority of the power circuit is warmed to 50–65ºC, and the rest of the circuit stays at around ambient temperature. No extreme hotspots were observed. The measurements also showed a good correlation with the calculations.

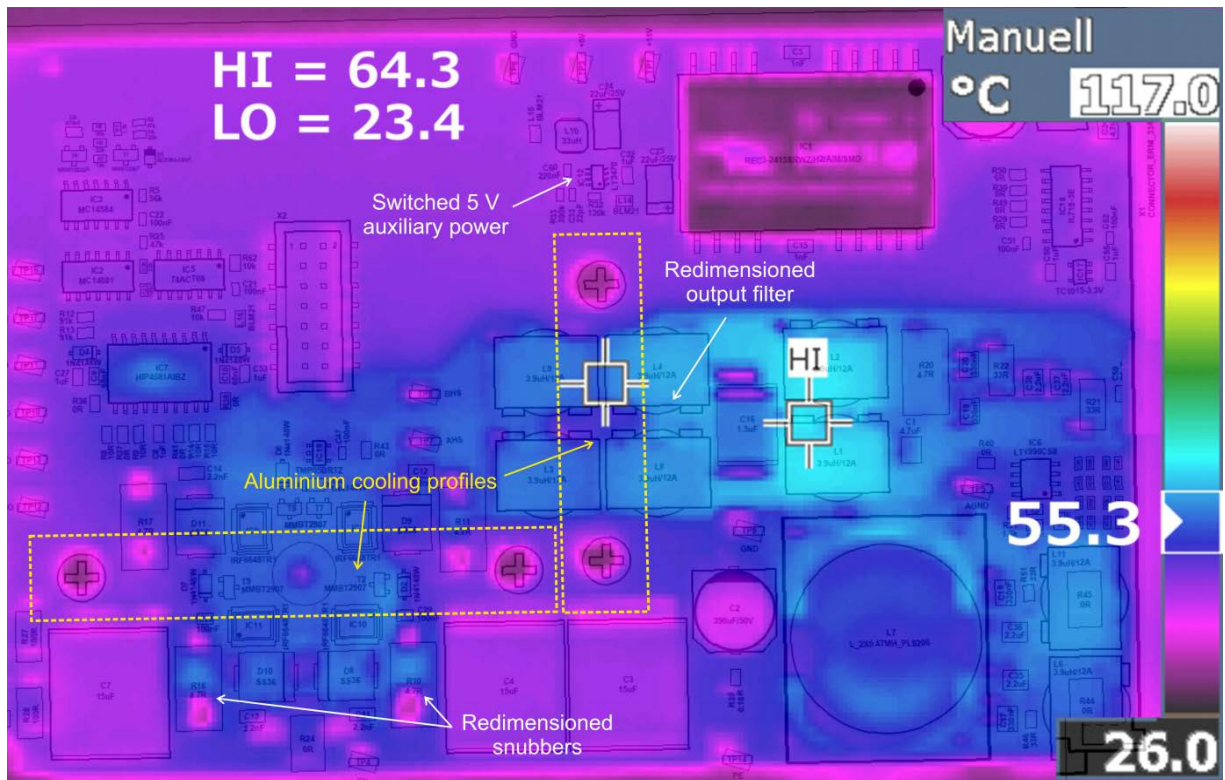

**Fig. 22:** Thermal image of the final design